# Designing a mobile game to thwarts malicious IT threats: A phishing threat avoidance perspective

Nalin A.G. Arachchilage[1]

*Australian Centre for Cyber Security (ACCS) School of Engineering and Information Technology UNSW, Canberra Australia*

Ali Tarhini[2]

*Department of Computer Science Brunel University London UK*

Steve Love[3]

*Digital Design Studio The Glasgow School of Art Glasgow, UK*

**Abstract**

*Phishing is an online identity theft, which aims to steal sensitive information such as username, password and online banking details from victims. To prevent this, phishing education needs to be considered. Game based education is becoming more and more popular. This paper introduces a mobile game prototype for the android platform based on a story, which simplifies and exaggerates real life. The elements of a game design framework for avoiding phishing attacks were used to address the game design issues and game design principles were used as a set of guidelines for structuring and presenting information. The overall mobile game design was aimed to enhance the user's avoidance behaviour through motivation to protect themselves against phishing threats. The prototype mobile game design was presented on MIT App Inventor Emulator.*

## 1. Introduction

The message "security is important" should be spread to all computer users. Computer users play a major role in helping to make cyberspace a safer place for everyone due to the growth of Internet technology. Internet technology is so ubiquitous today that it provides the backbone for modern living enabling ordinary people to shop, socialize, communicate, network, and be entertained all thorough their personal computers. As people's reliance on the Internet grows, so the possibility of hacking and other security breaches increases regularly [5, 28].

Security exploits can include malicious IT threats such as computer programs which can disturb the normal behaviour of computer systems (viruses), malicious software (malware), unsolicited e-mail (spam), monitoring software (spyware), attempting to make computer resources unavailable to its intended users (Distributed Denial-of-Service or DDoS attack), the art of human hacking (social engineering) and online identity theft (phishing). The motivation behind these attacks tends to be for either financial or social gain [30, 31, 43, 45]. For example, a DDoS attack could target a bank in order to break down their email server and the attacker can exhort a lump sum of money to give the email server back to the bank.

One such malicious IT threat that is particularly dangerous to computer users is phishing [14, 51]. Phishing, however, is a form of *semantic attack* [37] and well known as online identity theft, which aims to steal sensitive information such as username, password and online banking details from its victims. In phishing, victims get invited by scam emails to visit mimic websites. The attacker creates a mimic website which has the look-and-feel of the legitimate website. Innocent users are invited by sending emails to access to the mimic website and steal their money. Phishing attacks get more sophisticated daily as and when attackers learn new techniques and change their strategies accordingly [24].

Automated anti-phishing tools have been developed and used to alert users of potentially fraudulent emails and websites. For example, Calling ID Toolbar, Cloudmark Anti-Fraud Toolbar, EarthLink Toolbar, Firefox 2, eBay Toolbar, and Netcraft Anti-Phishing Toolbar. However, these tools are not entirely reliable in detecting phishing attacks [38, 39]. Even the best anti-phishing tools missed over 20 percent of phishing websites [49]. Ye et al. [48] and Dhamija and Tygar [13] have developed a prototype called "trusted paths" for the Mozilla web browser that is designed to help users verify that their browser has made a secure connection to a trusted website. However, these systems are still insufficient to combat phishing threats [5, 6, 7, 8, 9, 24, 25].

Security experts and phishing attackers are in a rat race today. On the one hand, security experts with the help of software designers and developers will continue to improve phishing and spam detection tools. Nevertheless, people are the weakest link in information security [11]. On the other hand, attackers will not hesitant to learn new techniques and change their strategies according to human defects to make a phishing attack successful [24, 25]. To prevent this, phishing education needs to be considered [6, 15, 24, 25, 35, 36, 38, 39, 52].

The aim of this research study is to design a game prototype as an educational tool to teach computer users how protect themselves against phishing attacks. Therefore, it asks the following questions: The first question is how does the system developer identify which issues the game needs to be addressed? Once the developer has identified the salient issues, they are faced with second question, what principles should be used to structure this information. The elements of a game design framework were used to address those mobile game design issues and the mobile game design principles were used as guidelines for structuring and presenting information in the game design context. Then a game prototype was designed for mobile android platform using MIT App Inventor Emulator.

The objectives are follows:
• To identify the elements should be addressed in the game design prototype for computer users to avoid themselves from phishing attacks.
• To design a storyboard for the game based on a story, which simplifies and exaggerates real life.
• To design the mobile game design prototype using MIT App Inventor Emulator for computer users to thwart phishing attacks.

## 2. Related Work

Previous research has indicated that technology alone is insufficient to ensure critical IT security measures. To date, there has been little work published on the human aspect of people performing security checks and preventing themselves from various attacks which are imperative to cope up with malicious IT threats such as phishing attacks ([1, 3, 5, 10, 19, 20, 27, 28, 31, 45, 46]. Many *Information Security* related discussions have finished with the conclusion of *if we could only remove the end-user from the system we would be able to make it secure* [18]. Where it is impossible to completely eliminate the end-user, for example in home computer use, the best possible approach for computer and information security is to educate the end-user in security prevention [29, 37]. Previous research has discovered well designed end-user security education can be effective [24, 25, 26, 38, 39]. This could be web-based training materials, contextual training, and embedded training to enhance users' ability to avoid phishing threats. One objective of the current research is to find effective ways to educate people to identify and avoid phishing attacks.

So, how does one educate computer users in order to prevent them from becoming victims of becoming victims of phishing threats? The study reported in this paper attempts to design and develop a mobile game as a tool to educate computer users against phishing attacks. This concept is grounded on the notion that not only computer games can provide education [50], but also games potentially offer a better natural learning environment, which motivates the user to keep engaging with it [2, 33]. In addition, game based education attracts and retains the user till the end of game by providing immediate feedback.

The most significant feature of a mobile environment is *mobility* itself such as mobility of the user, mobility of the device, and mobility of the service [32]. It enables users to be in contact while they are outside the reach of traditional communicational spaces. For example, a person can play a game on his mobile device while travelling on the bus or train, or waiting in a queue.

Some innovators strongly argue that desktop computers will disappear from the society while new handheld devices and their interfaces will turn into ubiquitous, pervasive, invisible, and embedded in the surrounding environment [40]. They also believe that those devices will be context-aware, attentive, and perceptive, sensing users' desires and providing feedback through ambient displays that glow, hum, change shape, or blow air. Furthermore, some researchers and technology experts predict advanced mobile devices that are wearable, or even implemented under the human skin. For example, the individual wireless sensors can use to track users entering premises [18].

Currently, there is a trend in games technology targeting dedicated handheld programming devices such as Nintendo DS and PlayStation Portable or PDAs and mobile phones [12, 22]. For example, touch-based interfaces introduced on Nintendo DS or iPhone/iPod changed computer based educational games to the emerging mobile-based platform. Those touch-based interfaces enable the player to interact with digital objects with in the gaming environment much easily than navigating through the keyboard. As a consequence of those emerging mobile technology, iPhone and Android devices and low-cost apps stores are awash with games.

## 3. Game Design Issue

The main focus of the proposed game design is to educate computer users to thwart phishing attacks. To answer this question issues drawn from a game design framework will be used to explore the principles needed for structuring the design of the game in the context of computer use. The game design framework describes individual computer users' behaviour of avoiding the threat of malicious information technologies such as phishing attacks [5]. The game design framework attempted to enhance individuals' phishing threats avoidance behaviour through their motivation to protect themselves from phishing attacks. The proposed game design framework is shown in Figure 1 [5]. The hypotheses (H) are described as follows:

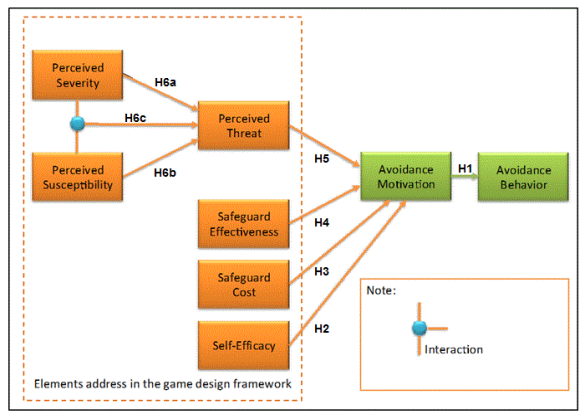

Figure. 1. A game design framework for avoiding phishing attacks [5]

H1. Avoidance motivation positively affects the avoidance behaviour.
H2. Self-efficacy positively affects avoidance motivation.
H3. Safeguard Cost negatively affects avoidance motivation.
H4. Safeguard Effectiveness positively affects avoidance motivation.
H5. Perceived Threat positively affects avoidance motivation.
H6a. Perceived Severity positively affects avoidance motivation.
H6b. Perceived Susceptibility positively affects avoidance motivation.
H6c. The combination of Perceived Severity and Perceived Severity positively affects avoidance motivation.

Consistent with the game design framework (Figure. 1), the users' IT threat avoidance behaviour is determined by avoidance motivation, which, in turn, is affected by perceived threat. Perceived threat is influenced by perceived severity and susceptibility as well as their combination. Users' avoidance motivation is also determined by the three constructs such as safeguard effectiveness, safeguard cost, and self-efficacy. In addition, the game design framework posits that perceived threat is influenced by the combination of perceived severity and susceptibility.

Whilst the game design framework forms the issues that the game design needs to be addressed, it should also indicate how to structure this information and present in a game context. Therefore, the game design attempts to develop threat perceptions, making individuals more motivated to avoid phishing attacks and use safeguarding measures. Finally, the elements of the game design framework were incorporated into mobile game prototype design to enhance individuals' phishing threats avoidance behaviour through their motivation to protect themselves from phishing attacks.

### 3.1 What to teach?

Possible phishing attacks can be identified in several ways, such as carefully looking at the website address, so called Universal Resource Locator (URL), signs and content of the web page, the lock icons and jargons of the webpage, the context of email message, and the general warning messages displayed in the website [15, 47]. Previous research has identified that existing anti-phishing techniques based on URLs are not robust enough for phishing detection [16, 21, 38, 39]. Garera et al. [16] strongly argued it is often possible to differentiate phishing websites from legitimate ones by carefully looking at the URL without having any knowledge of the content of the corresponding website, sings and symbols such as "VeriSign" signs or "Padlock" icons. Therefore, this research argues that teaching people not to fall for phishing through URLs is still infancy, but well-designed anti-phishing education based on URLs can contribute to stop users falling for phishing.

The objective of the anti-phishing mobile game design, reported in this paper, is to teach user how to identify phishing website addresses (URLs). Therefore, the game design should develop an awareness of identifying the features of website addresses. For example, legitimate websites usually do not have numbers at the beginning of their URLs such as http://81.153.192.106/.www.hsbc.co.uk.

### 3.2 Story and Mechanism

Funny stories are a great tonic for maintaining users' attraction in eLearning [41]. Storytelling techniques are used to grab attention, which also exaggerate interesting aspect of reality [6]. Stories can be based on personal experiences or famous tales or it could also be aimed to build a storyline that associates content units, inspire or reinforce.

The game is based on a scenario of a character of a small fish and 'his' teacher who live in a big pond.

The more appropriate, realistic and content relevant the story, there would be better the chances that it will trigger users. The main character of the game is the small fish, who wants to eat worms to become a big fish. The game player roll plays as a small fish. However, he should be careful of phishers those who try to trick him with fake worms. This represents phishing attacks by developing threat perception. Each worm is associated with a website address (URL), which appears in a dialog box. The game was designed total 10 URLs to randomly display including five good worms and five bad worms. The small fish's job is to eat all the real worms which associate legitimate website addresses and reject fake worms which associate with fake website addresses before the time is up. This attempts to develop the severity and susceptibility of the phishing threat in the game design.

The other character is the small fish's teacher, who is a matured and experienced fish in the pond. If the worm associated with the URL is suspicious and if it is difficult to identify, the small fish can go to 'his' teacher and request help. The teacher could help him by giving some tips on how to identify bad worms. For example, "website addresses associate with numbers in the front are generally scams," or "a company name followed by a hyphen in a URL is generally a scam". Whenever the small fish requests help from the teacher, the score will be reduced by certain amount (in this case by 100 seconds) as a payback for safeguard measure. This attempts to address the safeguard effectiveness and the cost needs to pay for the safeguard in the game design. The consequences of the player's actions are shown in Table 1.

The proposed game design randomly generates a worm associate with a URL each time. The URL could be either phishing or legitimate. When the user comes across the game from the beginning to the end, complexity of URLs is dramatically increased. The user is presented a worm associated with a different URL each time throughout the game. This helps the user to gain the conceptual knowledge of identifying URLs. Therefore self-efficacy of preventing from phishing attacks will be addressed in the game design as the user comes across through the game.

The proposed game design is based on story and presented to the player using digital objects. By creating attractive digital objects in the game design, not only immerse in an augmented physical environment, but also immerse into an augmented social environment. The overall game design aim is used to enhance individual users' avoidance behaviour through motivation to protect themselves against phishing attacks.

### 3.3 Technology

The content for the game, including URLs and training tips, are hard coded during the design and implementation. The App Inventor tool provided a great deal of flexibility and made easy to quick update the content in the mobile game prototype [44]. This was one of the main reasons to use MIT App Inventor emulator for developing the high-fidelity prototype. The current research employed the approach of URL classification used in Sheng et al. and Dhamija et al. studies [14, 38, 39]. Therefore, the game was designed total 10 URLs to randomly display including five good worms and five bad worms. The list of URLs is shown in Table 2.

Table1. Scoring scheme and consequences of the player's action

|  | **Good worm (associated with legitimate URL)** | **Bad worm (associated with phishing URL)** |
|---|---|---|
| **Player eats** | Correct, gain 10 points (each attempt = 1 point) | False negative, (each attempt loses 100 seconds out of 600 seconds) |
| **Player reject** | False positive, (each attempt loses 100 seconds out of 600 seconds) | Correct, gain 10 points (each attempt= 1 point) |

Table 2. List of URLs displayed in the game

| Game Focus | Real or Phishing | Examples | "Tips/ Training messages" from big fish |
|---|---|---|---|
| Appropriate URL | *Real* | *http://www.nationwide.co.uk/default.htm* | "URLs with well-known domain and correctly spelled are legitimate" |
| IP address URL | *Phishing* | *http://147.46.236.55/PayPal/* | "Don't trust URLs with all numbers |

| | | *login.html* | in the front" |
|---|---|---|---|
| Miss spelled URL | *Phishing* | *www.paypa1.com* | "Don't trust URLs with misspelled known websites" |
| Appropriate URL | *Real* | *www.smile.co.uk/* | "URLs with well-known domain and correctly spelled are legitimate" |
| Sub domain URL | *Phishing* | *www.argos.co.uk.myshop.com* | "Don't trust URLs with large host names that contained a part of a well-known web addresses" |
| Similar and deceptive domains | *Phishing* | *http://www.msn-verify.com/* | "Company name followed by a hyphen usually means, it's a scam website" |
| Appropriate URL | *Real* | *http://www.halifax.co.uk/aboutonline/home.asp* | "URLs with well-known domain and correctly spelled are legitimate" |
| Similar and deceptive domains | *Real* | *www.ebay-security.com* | "Companies don't use security related keywords in their domains" |
| Miss spelled URL | *Phishing* | *www.online.ll0ydstsb.co.uk* | "Don't trust URLs with misspelled known websites" |
| Appropriate URL | *Real* | *https://ibank.barclays.co.uk/* | "URL with 'https://' usually a legitimate website" |

In addition, attractive digital objects were integrated such as sounds and graphics with the game to better engage the user within the gaming environment. This included sounds effects to provide feedbacks on underwater background music and the player's actions on the selection of either good or bad worm. For example, a light water bubbling sound was played in the background throughout the game to feel that the user lives in the pond.

## 4. Game design principle

The above mentioned scenario based on the elements of game design framework (shown in Figure 1) should combine with a set of guidelines that focused on designing of the educational game [2, 33, 34]. This set of guidelines is known as game design principles, which describes how the user interacts with mobile game. Prensky [33, 34] has proposed that the mobile game can be described in terms of six structural elements. Those elements were used in the game design as guidelines for structuring and presenting information.

1. Rules: which organize the game. The story developed based on the elements of the game design framework describes the rules.
2. Goals and objective: which the players struggle to achieve. The user has the goal to solve the task. This is designed in the game to complete it from begin to end by eating real worms associated with legitimate URLs while avoiding fake worms associated with fake URLs.
3. Outcome and feedback: which measure the progress against the goals. The user gets the real time feedback on the current status in the game. For instance, when the user taps on a real worm, which is associated with legitimate URL, a real time audio feedback is played saying "WOW well done". Similarly, when the users taps on a fake worm, which is associated with fake URL, also a real time audio feedback is displayed saying "Oh Try again".
4. Conflict, competition, challenge, and opposition leading to players' excitement: which are realised through the opportunity to gain points against given lives of the player in the mobile game design.
5. Interaction: which is known as the social aspect in the game. This is accomplished by providing real time feedback, fantasy and rewords or gaining points in the game design. By creating attractive digital objects such as a big fish (in this case, the teacher), worms, pond in the game design, not only immerse in an augmented physical environment, but also immerse into an augmented social environment.
6. Representation or story: which exaggerates interesting aspects of reality. The representation is realized through the scenario or the story developed using digital objects such as the small fish (game player), 'his' teacher, and worms in the game design.

Furthermore, the work not only focuses on the design of a mobile game [23], but also it should focus on the usability of the game application [17]. Usability of the game design often attempts to recreate a typical playing environment, to emulate how a player would typically play the game in the real world while allowing game players to achieve their goals. For example, they should notice when something in the game changes such as decreasing the time period while intricate the combination of URLs when they move the digital objects in the game using keys in the touch stick or finger without any difficulty.

The game scenario and design principles (a set of guidelines) were combined together to focus on designing an educational mobile game for computer users to thwart phishing attacks. Therefore, this study attempted to design a storyboard for the game as the initial step, based on the game scenario and design principles.

## 5. Mobile Game Prototype

The research study attempted to design and develop a mobile game prototype for computer users to thwart phishing attacks based on the game design framework introduced by Arachchilage and Love [5]. To explore the viability of using a game to prevent from phishing attacks, a working prototype model was developed for a mobile telephone using MIT App Inventor Emulator (Fig. 3). Google app inventor is an application originally introduced by Google and now maintained by MIT (Massachusetts Institute of Technology), which is also a high level visual programming tool [44]. It uses a visual "blocks" language for creating mobile applications for android platform, which is easy, cost effective and less time consuming for developing prototypes compare to other traditional programming languages ("App Inventor Classic," [4]). Therefore, this study was employed MIT App Inventor Emulator as a visual block programming language to create a mobile game prototype model, which is shown in Figure 2.

The advantage of creating a workable mock-up enables designers to visually convey the envisioned interaction for a specific class of applications [42]. For example, mobile game mock-up can act as high-fidelity prototype versions of the application, thus designers can use to study the usability of overall application.

The player is given instructions before starting the game. Then the main menu of the mobile game mock-up is appeared along with underwater background sound effects. A light water bubbling sound is played in the background throughout the game to feel the user that he is in the pond. A URL is displayed with each worm where the worms are randomly generated.

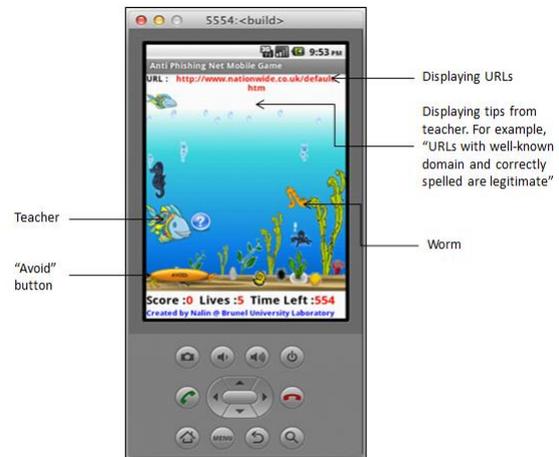

Figure 2. Main Menu of the game prototype design displayed on MIT App Inventor Emulator.

If the worm associated with URL is legitimate, then the user is expected to tap on the worm in order to increase the score. However, if the user fails to identify the legitimate URL, then remaining lives will be reduced by one point. On the other hand, if the worm associated with the URL is phishing, then the user is also expected to tap on "AVOID" button to reject the URL in order to increase the score. If the user fails to do this, then remaining lives will be reduced by one point. If the worm associated with the URL is suspicious and if it is difficult to identify, the user can tap on big fish (in this case, teacher fish) to request help. Then some relevant tips will be displayed just below the URL. For example, "website addresses associate with numbers in the front are generally scams. Whenever the user taps on the big fish, the time left will be reduced by 100 points (in this case 100 seconds).

Finally, the user will gain 10 points if all the given URLs were correcly identifed within 5 lives and 600 seconds to complete the game.

## 6. Conclusion and Future Work

This research focuses on design a mobile game as a tool to educate home computer users by creating awareness of phishing attacks. It addressed two questions: The first question is how does the systems developer identify which issues the game needs to be addressed? Then they are faced with second question, what principal should guide them to structure this information? We used a game designed framework in order to address those issues and mobile game design principles were used as a set of guidelines for structuring and presenting information in the game design context. The objective of our anti-phishing mobile game design was to teach the user how to identify phishing website addresses (URLs) which is one of many ways to identify a phishing attack. As future research we will empirically evaluate this mobile game prototype to check whether or not it teaches to protect computer users from phishing attacks. In addition, future research can be addressed on designing a mobile game to teach the other areas such as signs and content of the web page, the lock icons and jargons of the webpage, the context of the email message, and the general warning messages displayed in the website. The overall game design was targeted to enhance avoidance behaviour through motivation to protect home computer users from phishing attacks. We believe by educating home computer users against malicious IT threats, would make a considerable contribution to enable the cyberspace as a secure environment.

## 10. References

[1] Alsharnouby, M., Alaca, F., & Chiasson, S. (2015). Why phishing still works: user strategies for combating phishing attacks. International Journal of Human-Computer Studies.

[2] Amory, A., Seagram, R., 2003. Educational game models: conceptualization and evaluation: the practice of higher education. South Afr. J. High. Educ. 17, 206.

[3] Anderson, C.L., Agarwal, R., 2006. Practicing safe computing: Message framing, self-view, and home computer user security behavior intentions, in: Proceedings of the 27th International Conference on Information Systems.

[4] App Inventor Classic [WWW Document], n.d. URL http://appinventor.mit.edu/explore/classic.html (accessed 2.15.14).

[5] Arachchilage, N.A.G., Love, S., 2013. A game design framework for avoiding phishing attacks. Comput. Hum. Behav. 29, 706–714.

[6] Arachchilage, N. A. G., & Love, S. 2014. Security awareness of computer users: A phishing threat avoidance perspective. *Computers in Human Behavior*, *38*, 304-312.

[7] Arachchilage, N. A. G., & Cole, M. 2011. Design a mobile game for home computer users to prevent from "phishing attacks". In *Information Society (i-Society), 2011 International Conference on* (pp. 485-489). IEEE.

[8] Arachchilage, N. A. G., Love, S., & Scott, M. 2012. Designing a Mobile Game to Teach Conceptual Knowledge of Avoiding "Phishing Attacks". *International Journal for e-Learning Security*, *2*(2), 127-132.

[9] Arachchilage, N. A.G., Flechais, I., & Beznosov, K.2014. A Game Storyboard Design for Avoiding Phishing Attacks. Symposium On Usable Privacy and Security (SOUPS), 2.

[10] Aytes, K., Connolly, T., 2005. Computer security and risky computing practices: A rational choice perspective. Adv. Top. End User Comput. 4, 257.

[11] CNN, 2005. A convicted hacker debunks some myths.

[12] Denk, M., Weber, M., Belfin, R., 2007. Mobile learning–challenges and potentials. Int. J. Mob. Learn. Organ. 1, 122–139.

[13] Dhamija, R., Tygar, J.D., 2005. The battle against phishing: Dynamic security skins. Presented at the ACM International Conference Proceeding Series, pp. 77–88.

[14] Dhamija, R., Tygar, J.D., Hearst, M., 2006. Why phishing works. Presented at the Proceedings of the SIGCHI conference on Human Factors in computing systems, ACM, pp. 581–590.

[15] Downs, J.S., Holbrook, M., Cranor, L.F., 2007. Behavioral response to phishing risk, in: Proceedings of the Anti-Phishing Working Groups. Presented at the 2nd Annual eCrime researchers summit, ACM, Pittsburgh, Pennsylvania, USA, pp. 37–44.

[16] Garera, S., Provos, N., Chew, M., Rubin, A.D., 2007. A framework for detection and measurement of phishing attacks. Presented at the Proceedings of the 2007 ACM workshop on Recurring malcode, ACM, 1314391, pp. 1–8.

[17] Gong, J., Tarasewich, P., 2004. Guidelines for handheld mobile device interface design, in: Proceedings of DSI 2004 Annual Meeting. Citeseer, pp. 3751–3756.

[18] Görling, S., 2006. The myth of user education, in: Virus Bulletin Conference. p. 13.

[19] Harrington, S., Anderson, C.L., Agarwal, R., 2006. Practicing Safe Computing: Message Framing, Self-View, and Home Computer User Security Behavior Intentions., in: ICIS. p. 93.

[20] Ion, I., Reeder, R., & Consolvo, S. (2015, July). "... no one can hack my mind": Comparing Expert and Non-Expert Security Practices. In Symposium on Usable Privacy and Security (SOUPS).

[21] Kirlappos, I., Sasse, M.A., 2012. Security Education against Phishing: A Modest Proposal for a Major Rethink. IEEE Secur. Priv. Mag. 10, 24–32.

[22] Klopfer, E., 2008. Augmented learning: Research and design of mobile educational games. MIT Press.

[23] Korhonen, H., Koivisto, E.M., 2006. Playability heuristics for mobile games, in: Proceedings of the 8th Conference on Human-Computer Interaction with Mobile Devices and Services. ACM, pp. 9–16.

[24] Kumaraguru, P., Rhee, Y., Sheng, S., Hasan, S., Acquisti, A., Cranor, L.F., Hong, J., 2007a. Getting users to pay attention to anti-phishing education: evaluation of retention and transfer, in: Proceedings of the Anti-Phishing Working Groups 2nd Annual eCrime Researchers Summit. ACM, pp. 70–81.

[25] Kumaraguru, P., Rhee, Y., Sheng, S., Hasan, S., Acquisti, A., Cranor, L.F., Hong, J., 2007b. Getting users to pay attention to anti-phishing education: evaluation of retention and transfer. Presented at the Proceedings of the anti-phishing working groups 2nd annual eCrime researchers summit, ACM, pp. 70–81.

[26] Le Compte, A., Elizondo, D., & Watson, T. (2015, May). A renewed approach to serious games for cyber security. In Cyber Conflict: Architectures in Cyberspace (CyCon), 2015 7th International Conference on (pp. 203-216). IEEE.

[27] Liang, H., Xue, Y., 2009. Avoidance of Information Technology Threats: A Theoretical Perspective. MIS Q. 33.

[28] Liang, H., Xue, Y., 2010. Understanding Security Behaviors in Personal Computer Usage: A Threat Avoidance Perspective. J. Assoc. Inf. Syst. 11.

[29] Mitnick, K.D., Simon, W.L., 2001. The art of deception: Controlling the human element of security. John Wiley & Sons.

[30] Ng, B.-Y., Kankanhalli, A., Xu, Y.C., 2009. Studying users' computer security behavior: A health belief perspective. Decis. Support Syst. 46, 815–825.

[31] Ng, B.-Y., Rahim, M., 2005. A Socio-Behavioral Study of Home Computer Users' Intention to Practice Security., in: PACIS. p. 20.

[32] Parsons, D., Ryu, H., Cranshaw, M., 2006. A study of design requirements for mobile learning environments, in: Advanced Learning Technologies, 2006. Sixth International Conference on. pp. 96–100.

[33] Prensky, M., 2003. Digital game-based learning. Comput. Entertain. CIE 1, 21–21.

[34] Prensky, M., 2005. Computer games and learning: Digital game-based learning. Handb. Comput. Game Stud. 18, 97–122.

[35] Richmond, R., 2006. Hackers set up attacks on home PCs, financial firms: study. September.

[36] Robila, S.A., Ragucci, J.W., 2006. Don't be a phish: steps in user education, in: ACM SIGCSE Bulletin. ACM, pp. 237–241.

[37] Schneier, B., 2000. Semantic attacks: The third wave of network attacks. Crypto-Gram Newsl. 14.

[38] Sheng, S., Magnien, B., Kumaraguru, P., Acquisti, A., Cranor, L.F., Hong, J., Nunge, E., 2007. Anti-phishing phil: the design and evaluation of a game that teaches people not to fall for phish, in: Proceedings of the 3rd Symposium on Usable Privacy and Security. ACM, pp. 88–99.

[39] Sheng, S., Magnien, B., Kumaraguru, P., Acquisti, A., Cranor, L.F., Hong, J., Nunge, E., 2007. Anti-Phishing Phil: The Design and Evaluation of a Game That Teaches People Not to Fall for Phish. Presented at the Symposium On Usable Privacy and Security, ACM.

[40] Shneiderman, B., 1987. Designing the user interface: supplemental materials.

[41] Taran, C., 2005. Motivation techniques in eLearning, in: Advanced Learning Technologies, 2005. ICALT 2005. Fifth IEEE International Conference on. IEEE, pp. 617–619.

[42] Truong, K.N., Hayes, G.R., Abowd, G.D., 2006. Storyboarding: an empirical determination of best practices and effective guidelines, in: Proceedings of the 6th Conference on Designing Interactive Systems. ACM, pp. 12–21.

[43] Warkentin, M., Johnston, A.C., 2006. IT security governance and centralized security controls. Enterp. Inf. Assur. Syst. Secur. Manag. Tech. Issues 16–24.

[44] Wolber, D., 2011. App inventor and real-world motivation, in: Proceedings of the 42nd ACM Technical Symposium on Computer Science Education. ACM, pp. 601–606.

[45] Woon, I., Tan, G.-W., Low, R., 2005. A protection motivation theory approach to home wireless security.

[46] Workman, M., Bommer, W.H., Straub, D., 2008. Security lapses and the omission of information security measures: A threat control model and empirical test. Comput. Hum. Behav. 24, 2799–2816.

[47] Wu, M., Miller, R.C., Garfinkel, S.L., 2006. Do security toolbars actually prevent phishing attacks? Presented at the Proceedings of the SIGCHI conference on Human Factors in computing systems, ACM, pp. 601–610.

[48] Ye, Z.E., Smith, S., Anthony, D., 2005. Trusted paths for browsers. ACM Trans. Inf. Syst. Secur. TISSEC 8, 153–186.

[49] Zhang, Y., Hong, J., Cranor, L.F., 2007. Cantina: a content-based approach to detecting phishing web sites. Presented at the Proceedings of the 16th international conference on World Wide Web, ACM, 1242659, pp. 639–648.

[50] E. M. Raybourn and A. Waern, "Social Learning Through Gaming", Proceedings of CHI 2004, Vienna, Austria, pp. 1733-1734, 2004.

[51] Dadkhah, M., Tarhini, A., Lyashenko, V. & Jazi, M. D. (2015). Hiring Editorial Member for Receiving Papers from Authors. Mediterranean Journal of Social Sciences, 6(4), 11-12.

[52] El-Masri, M., Tarhini, A., Hassouna, M., & Elyas, T. (2015, May). A design science approach to Gamify education: From games to platforms. In Twenty-Third European Conference on Information Systems (ECIS).